# Inferring correlations associated to causal interactions in brain signals using autoregressive models

Víctor J. López-Madrona[1,*], Fernanda S. Matias[2,3], Claudio R. Mirasso[4], Santiago Canals[1], Ernesto Pereda[5,6,*]

[1] Instituto de Neurociencias, CSIC-UMH, Sant Joan d'Alacant 03550, Spain

[2] Cognitive Neuroimaging Unit, Commissariat à l'Energie Atomique (CEA), INSERM U992, NeuroSpin Center, 91191 Gif-sur-Yvete, France

[3] Instituto de Física, Universidade Federal de Alagoas, 57072-970 Maceió, Alagoas, Brazil

[4] Instituto de Física Interdisciplinar y Sistemas Complejos, IFISC (UIB-CSIC), Campus Universitat de les Illes Balears E-07122, Palma de Mallorca, Spain

[5] Departamento de Ingeniería Industrial, Escuela Superior de Ingeniería y Tecnología, IUNE, Universidad de La Laguna, Tenerife 38205, Spain

[6] Laboratory of Cognitive and Computational Neuroscience, CTB, UPM, Madrid, Spain

**\* Corresponding Authors**:

Víctor J. López-Madrona (v.lopez@umh.es)

Ernesto Pereda (eperdepa@ull.edu.es)

**Abstract**

The specific connectivity of a neuronal network is reflected in the dynamics of the signals recorded on its nodes. The analysis of how the activity in one node predicts the behaviour of another gives the directionality in their relationship. However, each node is composed of many different elements which define the properties of the links. For instance, excitatory and inhibitory neuronal subtypes determine the functionality of the connection. Classic indexes such as the Granger causality (GC) quantifies these interactions, but they do not infer into the mechanism behind them. Here, we introduce an extension of the well-known GC that analyses the correlation associated to the specific influence that a transmitter node has over the receiver. This way, the G-causal link has a positive or negative effect if the predicted activity follows directly or inversely, respectively, the dynamics of the sender. The method is validated in a neuronal population model, testing the paradigm that excitatory and inhibitory neurons have a differential effect in the connectivity. Our approach correctly infers the positive or negative coupling produced by different types of neurons. Our results suggest that the proposed approach provides additional information on the characterization of G-causal connections, which is potentially relevant when it comes to understanding interactions in the brain circuits.

# 1- Introduction

The concept of connectivity in neuroscience is closely related to two main subjects: anatomical and functional networks. The first one refers to the relatively static physical links between neurons, the so called wiring diagram of the brain network. It is mainly composed by neuronal axons organized in fibre tracts and, while it changes in long-term (ontogenetic) time scales (during development, maturation and aging of organisms), it is considered stable in most studies in healthy adults investigating fast brain dynamics (from milliseconds to hours to days or even months). Structural networks can be quantified by histological techniques and microscopy, then providing directionality in the connections or a directed graph, or estimated by non-invasive imaging techniques such as diffusion tensor imaging, producing undirected graphs[1,2]. This is the realm of functional connectivity, which studies the statistical interdependencies between the activities of different areas and is commonly measured with indexes like correlation or coherence. As such, functional connectivity produces un-directed functional graphs. There is yet another definition for connectivity, one concerned with functional interactions but in a directed manner, named, effective connectivity. This connectivity considers the effect of one node over another, establishing causality in the interaction. In recordings of time series, this connectivity is commonly estimated under the assumption that the sender (cause) must precede the receiver (effect) in time[3]. Techniques such as the Local Field Potential (LFP) and the electroencephalography (EEG), measure the postsynaptic potential of hundreds of neurons, reflecting the average of their electrical activity. The synchronized spiking of cell assemblies gives rise to oscillations in extracellular recordings[4–6] (for example, theta and gamma rhythms in the hippocampus[7,8], with distinct proposed roles). Understanding the rhythms' properties, couplings, and the mechanisms that generate them can have direct clinical benefits, such as in the diagnosis and treatment of patients with epilepsy and other neurological disorders[9,10].

Studies of the causal relationships between time-varying signals were initiated by Wiener[11], who hypothesized that if the forecasting of a signal can be improved by incorporating information from the past values of another signal, then there is a causal influence of the latter on the former. This idea was later implemented by Granger through the use of autoregressive (AR) models, a methodology named after him as the Granger causality[12,13] (GC). This method has become widely applied in many fields since then. In neuroscience, it has boosted the characterization of brain circuits, revealing remarkable findings as the different pathways for alpha and gamma activities in the visual cortex[14], the leading role of the gamma waves in the right over the left hippocampus[15] or the bidirectional communication between theta rhythms in the medial septum and the hippocampus[16,17].

Despite its extensive use, there are some limitations and misunderstandings of GC[18] which, although partially solved in modern approaches[19–21], have generated uncertainty about its efficacy. Specifically, two main considerations have been recently pointed out as a limit to the application and interpretation of the GC[18]. First, the GC may be severely biased and of high variance when estimating separated and independent full and partial AR models. Even if the full model is a finite model order regression, the reduced model may be an infinite order process or have a *moving average* component, being a vector-autoregressive moving average (VARMA) process. These processes are poorly approximated by AR models and result in biased results of connectivity. Nevertheless, this problem has been previously acknowledge[22,23] and solved by methods that do not require reduced models[21,23,24]. Other approaches like state-space (SS) processes[25,26] characterize an observed time series by an unobserved state driven by a stochastic process, and are equivalent to VARMA processes[27,28], solving the problem of the moving average component[25,26]. Second, the GC is influenced by the dynamics of the sender and the channel, but not by the receiver[18]. This limitation has been argued to be a misunderstanding of the concepts of *causality* and *functional connectivity*. Methods like GC stablish a statistical relation based on observed responses and are measurements of *directed functional connectivity*. On the other hand, the identification of the mechanisms that produce an effect in a neural system are referred as *causality* or *effective connectivity* and they cannot be achieved solely by indexes like GC and require structural information, included in other methodologies as the dynamic causal modelling (DCM)[3,19,29,30]. While it is true that it should be used carefully, the GC is a reliable methodology to infer brain connectivity and new approaches are continuously being developed to overcome its limitations[19–21].

Several other methodologies have been developed to measure the directed functional connectivity in brain networks, considering different features of the time series[31,32]. For instance, Baccalá and Sameshima defined the direct coherence[33] (DC) and the partial directed coherence[34] (PDC), two indices based on AR models which allowed a spectral characterization of the connectivity, where the former measures the power that spreads from one structure to another (either directly or indirectly) and the latter represents only the direct pairwise interaction. Note that the distinction between direct (partialized) and indirect is made only with the observed variables, that is, with the signals included in the model. As most neuroscience data are nonstationary and the statistical connectivity may vary along the time, Dhamala and collaborators proposed a spectral time varying GC based on nonparametric statistics[35]. Instead of solving an AR model, they derived the frequency components directly through Fourier and wavelet transform of the data, bypassing the parametric data modelling. Other methodologies to estimate the directionality without AR models are based on the information theory, as the transfer entropy (TE)[36] or the phase transfer entropy (PhTE)[37]. The main advantage is the lack of linear modelling allowing them to detect nonlinearities in the connectivity. On the contrary, they do not have a spectral representation and require long time series to compute the indices.

Brain computations are realized in a densely but sparsely connected network of excitatory and inhibitory neurons. Integration of synaptic inputs and firing of excitatory neurons are often equated with information encoding and transmission, and respectively, with the inhibitory activity setting the pace of communication by coordinating the nodes and establishing windows of opportunity for the transmission[38–41]. Current connectivity indexes quantify the degree of interaction and directionality between brain signals, but they cannot elucidate their functional role without precise knowledge of the underlying wiring diagrams. Here, we propose going one step further by characterizing whether the activity of the receiver follows directly or inversely the dynamics of the sender. Therefore, we define the relationship as positive or negative GC respectively, by using selected elements of AR models. We validate our method in virtual networks, where brain signals are simulated as the extracellular currents, similar to the experimentally measured LFP. Under the theoretical assumption that excitatory and inhibitory activities reflect direct and inverse relationships between populations, we test the efficiency of our approach in different scenarios, combining solely excitatory and inhibitory projections with mixing activity of both sources.

In the first section of this paper, we review the concept of GC and present the theory behind our approach. We then test and validate our method with time series generated numerically from neuronal models of anatomical motifs in which the structural connectivity is known. Finally, we discuss its advantages and limitations and suggest future lines of research in this framework.

## 2- Methods

### 2.1- Granger causality

Given two time series, $X(t)$ and $Y(t)$, it is possible to construct an AR model for each signal in which each time point is the linear combination of their $p$ past samples plus a residual:

$$X(t) = \sum_{k=1}^{p} A_{1k} X(t-k) + \varepsilon_1(t)$$

$$Y(t) = \sum_{k=1}^{p} A_{2k} Y(t-k) + \mu_1(t) \qquad (1)$$

The coefficients in the $A$ matrix represent the weights of the contributions of past values $X$ and $Y$ in the prediction, and $\varepsilon$ and $\mu$ are stationary zero mean white noise processes which represent the prediction error. The model order $p$ can be estimated following different criteria, with the Akaike Information Criterion[42] (AIC) and the Bayesian Information Criterion[43] (BIC) being two of the most common choices:

$$AIC = \log(\det(\Sigma)) + \frac{2pn^2}{T}$$

$$BIC = \log(\det(\Sigma)) + \frac{\ln(T) pn^2}{T} \qquad (2)$$

where $T$ is the sample size, $n$ is the number of variables and $\Sigma$ is the covariance noise matrix defined as:

$$\Sigma = \begin{pmatrix} var(\varepsilon) & cov(\varepsilon, \mu) \\ cov(\mu, \varepsilon) & var(\mu) \end{pmatrix} \qquad (3)$$

If we add to the equation 1 the information that one of the variables has on the prediction of the other, we obtain a bivariate AR model:

$$X(t) = \sum_{k=1}^{p} A_{11k}X(t-k) + A_{12k}Y(t-k) + \varepsilon_2(t)$$

$$Y(t) = \sum_{k=1}^{p} A_{21k}X(t-k) + A_{22k}Y(t-k) + \mu_2(t)$$

(4)

If the variance of the error of the bivariate model $var(\mu_2)$ is smaller than that of the univariate one $var(\mu_1)$, then the past of $Y(t)$ improves the prediction of $X(t)$; in other words, $Y(t)$ Granger-causes (G-causes) $X(t)$. An advantage of the GC is that it can be extended for multivariable conditions and estimates direct and indirect links between the included nodes. For example, let us suppose that a third variable $Z(t)$ exists and the connectivity of the network is $Z \rightarrow Y \rightarrow X$. Then, using the bivariate model in equation 4 with the variables $X(t)$ and $Z(t)$, an interaction from $Z(t)$ to $X(t)$ would be detected, even if it is mediated by $Y(t)$. This issue is solved by including the three variables in a new model:

$$X(t) = \sum_{k=1}^{p} A_{11k}X(t-k) + \sum_{k=1}^{p} A_{12k}Y(t-k) + \sum_{k=1}^{p} A_{13k}Z(t-k) + \varepsilon_3(t)$$

$$Y(t) = \sum_{k=1}^{p} A_{21k}X(t-k) + \sum_{k=1}^{p} A_{22k}Y(t-k) + \sum_{k=1}^{p} A_{23k}Z(t-k) + \mu_3(t) \quad (5)$$

$$Z(t) = \sum_{k=1}^{p} A_{31k}X(t-k) + \sum_{k=1}^{p} A_{32k}Y(t-k) + \sum_{k=1}^{p} A_{33k}Z(t-k) + \eta_3(t)$$

Comparing the residuals in both the bivariate and multivariate conditions ($\varepsilon_2$ or $\varepsilon_3$), the variance of $\varepsilon_3$ would be the same as that of $\varepsilon_2$, as the past values of $Z(t)$ do not improve the forecasting because all the indirect contributions over $X(t)$ are already included in $\varepsilon_2$, only using the information in $Y(t)$.

While this procedure represents the main theory behind the detection of GC, the combination of reduced and full regressions (i.e. equations 4 and 5 respectively) could introduce a bias in the estimation of the GC[18,22,23]. This issue lies in the definition of the model order of the AR model. While the full AR model may have a finite model order, the same model order is considered when it is decomposed into the reduced regression. However, these processes are generally of infinite order, have a moving average component (VARMA processes) and are poorly modeled using the AR models. One way to overcome this limitation is by the characterization of the model as a SS process instead of an AR process. This way, the observed multiple time series depend upon a possibly unobserved state by an observation matrix and a state transition matrix[25–28] which may be a representation of a VARMA processes with a equivalence between the coefficient $A$ matrix and the state transition matrix[27,28].

Note that the estimation of a full AR model is usually limited to the number of nodes with available information in the network. In neuroscience, the variables of the AR model represent different brain units or regions, where the time-series are the recordings (e.g. EEG, fMRI, MEG) of these areas. However, the measurement of all the elements of the circuit is generally unfeasible and the inferred AR model is constrained by the data. These unknown or unrecorded variables introduce a confounding effect in the AR that may drastically affect the estimation of the functional connectivity[41,44].

Several algorithms and methods can be used to compute GC, including different statistical analyses to check both the AR model and the result of the analysis[23,45,46]. In this paper, we used the Matlab toolbox recently proposed by Barnett and Seth[23], which overcomes some limitations of previous toolboxes[47], including the estimation of GC with a single full regression.

**2.2- Determining positive and negative Granger causality**

The second step consists in identifying whether there is a positive or negative interaction between the sender and the receiver. Let us consider the paradigm where $X(t)$ and $Y(t)$ are the LFP recordings of two different populations. If we assume that there is an excitatory link from $Y(t)$ to $X(t)$, an increase of the spiking activity in population $Y(t)$, reflected by an increase of the amplitude of $Y(t)$, would lead to a variation in the future values of $X(t)$. In return, in the case of an inhibitory connection, assuming the same $Y(t)$, we can expect an increase of negative currents in the population $X(t)$ or, in other terms, a negative correlation between $Y(t)$ and $X(t)$. As these changes explain only a fraction of the total variance of both LFPs, the type of interaction cannot be identified straightforwardly. The idea behind our approach is that this information can be extracted from the coefficients $A_{ij}$ of the AR model. Thus, we selected those $A$ components that assess the influence of $Y(t)$ to $X(t)$ ($A_{12k}, k = 1, ...p$). If the average of all $A_{ij}$ coefficients is positive, we assume the input from $Y(t)$ to be excitatory. Otherwise, if it is negative, then the input from $Y(t)$ is regarded as inhibitory.

Estimating these coefficients is usually not simple. Even if a single full regression with the correct model order is considered in the analysis, the recordings are not always clean enough to extract the interactions intra- and inter-populations, and sometimes the time series are too short compared to the optimal model order. Moreover, the algorithms used to solve the equations of the AR models tend to assign a value to every $A_{ij}$ coefficient – even to those not improving the forecasting. Thus, we propose to estimate which components of the $A$ matrix do not significantly contribute to the AR model by performing multiple versions of the full AR model in equation 5. In each version we set to zero specific coefficients of the matrix $A$ and test as many models as possible combinations of $A$ can be realized with the coefficients forced to zero. Note that we are not building a reduced model but estimating which components of the full regression model are not contributing to the prediction. Once those parameters that are not useful have been identified, a single full AR model is used to infer the functional connectivity. The optimum zero constraints are those belonging to the AR model that minimizes a modified version of the information criteria. The previously defined AIC and BIC terms are redefined by[28]:

$$AIC' = \log(\sigma^2) + \frac{2m}{T}$$
$$BIC = \log(\sigma^2) + \frac{\ln(T)m}{T}$$
(6)

where $\sigma^2$ is the sum of the estimated residuals (i.e. $\varepsilon_2, \mu_2$ ...) divided by the sample size $T$ and $m$ is the number of estimated parameters. For the original model, the number of parameters for each equation is the model order $p$ multiplied by the number of variables in the model.

As it was said before, in each version of the AR model certain $A$ coefficients are set to zero. For example, in a one model only the coefficient $A_{ijp}$ is set to zero, in a second model only the coefficient $A_{ijp-1}$ is set to zero and in a third one, both $A_{ijp}$ and $A_{ijp-1}$ are set to zero. Therefore, the total number of matrices to test with possible zero constraints is $2^{(K^2 p)}$, where $K$ is the number of variables in the model. As this value is usually very large, the computational cost for every case can be impracticable. Several strategies have been proposed to overcome this limitation, and we suggest to apply a combination of bottom-up and top-down strategies in order to reduce the number of matrices to test, as it has been previously described in [28].

**2.2.1- Bottom-up strategy**

In general, the order of the selected model should be long enough to capture the slowest relationship between the variables we consider. However, not all lags provide useful information. For example, if we select a model order that is higher than that estimated by the information criteria, we are taking more coefficients than necessary to compute the AR model. Although an increase in the model order yields a better fitting of the data set, since more coefficients can be adjusted to reduce the residue, this improvement is not significant for orders higher than those estimated by the information criteria. In these cases, the algorithms tend to fit the noise instead of the real interactions.

Since the number of coefficients that are necessary to model every link in a system can be different for each interaction, some type of regularization, which does not include more coefficients than necessary, would be desired. The idea behind this method is to compute the minimum order for the contribution of one time series over another. Instead of using the whole model, the equation of each variable is analysed separately. That is, in order to determine the constraints of equation 5, we consider the model:

$$X(t) = \sum_{k=1}^{p_{xx}} A_{11k} X(t-k) + \varepsilon(t) \qquad (7)$$

where $p_{xx}$ is the optimal model, which minimizes the selected criterion. To find the value of $p_{xx}$, where $p_{xx} = p, p-1, \ldots, 0$, the information criterion in equation 6 is computed for the model in equation 7. In each iteration, the model order is decreased by 1 and if the information criterion is lower than the one estimated in the previous iteration (i.e. the new iteration minimizes it), the model order is reduced keeping the same variance for the residue. This step is repeated until a new iteration does not improve the information criterion (or $p_{xx} = 0$). In the next step, $p_{xx}$ is fixed to its optimum value and a new variable is added into the equation, fitting a new model:

$$X(t) = \sum_{k=1}^{p_{xx}} A_{11k}X(t-k) + \sum_{k=1}^{p_{xy}} A_{12k}Y(t-k) + \varepsilon(t) \qquad (8)$$

After these steps, a new model order is computed for the participation of one variable to each equation separately, and the coefficients beyond these values are set to zero.

### 2.2.2- Top-down strategy

Again, we use each variable separately for this method. After setting the excess of coefficients to zero with the previous strategy, the contribution of each remaining value is tested. Initially, we compute the new information criterion value. Then, starting from the furthest nonzero lag, each coefficient is set to zero and the model is estimated again. If the new information criterion is lower, we update the model keeping that zero constraint. As an example, let us consider a system with variables $X(t)$, $Y(t)$ and $Z(t)$. The equation of one of the variables would be:

$$X(t) = \sum_{k=1}^{p_{xx}} A_{11k}X(t-k) + \sum_{k=1}^{p_{xy}} A_{12k}Y(t-k) + \sum_{k=1}^{p_{xz}} A_{13k}Z(t-k) + \varepsilon(t) \qquad (9)$$

The first coefficient to test is $A_{11p_{xx}}$, then $A_{11p_{xx}-1}$ and so on. If when setting one coefficient to zero the information criterion minimizes, that constraint is kept in further iterations. Otherwise, it remains unchanged and the next iteration is computed.

These strategies allow us to test every single value at once, considerably reducing the computational cost. The total number of required operations is a multiple of $K^2p$, and increases linearly with the number of variables and the model order, avoiding the exponential growth when all matrices are tested. Alternative methods can also be applied to determine the constraints of the $A$ coefficients, each with specific advantages and disadvantages (see [28] for details).

### 2.2.3- Estimating positive and negative Granger causality ratio

Once the linear constraints have been established and the corresponding $A$ coefficients eliminated, we estimate the sign of the GC (sGC) as the following ratio:

$$sGC_{ij} = \frac{\sum_{k=1}^{p}(A_{ijk} > 0)^2}{A_{max_{ij}}} - \frac{\sum_{k=1}^{p}(A_{ijk} < 0)^2}{A_{max_{ij}}} \quad (10)$$

where the denominator:

$$A_{max_{ij}} = \max\left(\sum_{k=1}^{p}(A_{ijk} > 0)^2, \sum_{k=1}^{p}(A_{ijk} < 0)^2\right) \quad (11)$$

is defined to normalize the index so that $sGC_{ij} \in [-1,1]$.

In our specific paradigm, a $sGC_{ij} \sim -1$ ($sGC_{ij} \sim 1$) indicates a dominant inhibitory (excitatory) connection in the link, whereas $sGC_{ij} = 0$ on a significant link (as assessed by the traditional GC) suggests a balance between excitation and inhibition.

## 2.3- Statistical significance

To assess the statistical significance of the defined index, the whole time-series were divided into time windows of the same length, being the main condition that the variables within these windows are *covariance stationary* [48] (i.e., they have constant mean and variance). If the data is acquired in an event-related task, i.e., the response to a stimulus is recorded, the fast changes of the signal may exhibit nonstationary epochs in each repetition. An adaptive multivariate AR model can be employed in those cases[49], where a single model is achieved based on the entire data set. Furthermore, the interaction between brain regions is time-varying. This means that a single model cannot reproduce the behaviour of the network. New algorithms involving the estimation of time-variant AR models have been developed to overcome that limitation and contribute to the study of the dynamics in brain connectivity[50]. The sGC was assessed as the average of the sGC over all windows. The window length was fixed to 5 seconds in this work.

To estimate the statistical significance associated to the sGC, we used a surrogate data analysis ($N = 2000$ in this work) by block-resampling, where each signal was cut at a random time point and the blocks were permuted. The size of the blocks was the same as that of the segments used for the estimation of the sGC, so that the difference between the surrogates and the original data, if any, could not be attributed to the different number of samples in each case. The block shuffling procedure has the advantage, unlike the phase reshuffling one, to preserve the possible relationship among the phases within each time series, which is known to be a confound when using surrogate data to estimate the existence of correlation between two data sets[51]. Moreover, this methodology outperforms other techniques based on linear Fourier approaches, as it preserves the irregular fluctuations of the signal (i.e. short transients of non-stationarity)[52]. In this way, the temporal interactions were broken, minimizing the distortion of the original dynamics. Then, the sGC for each surrogate data was computed following these two criteria. First, no prior GC analysis or strategies to reduce the number of A coefficients were applied. These methodologies would include supplementary statistical tests of connectivity, instead of limiting the analysis to the significance of the sGC value. Secondly, the denominator in equation 10 was the same as in the original time-series. Thus, the surrogate sGC value was normalized respecting the original signal, sharing the same scale. The significance was then assessed, as usual, by comparing the value obtained from the original data with the distribution of the values obtained from the surrogates. For this, the normality of the distribution was firstly confirmed using the Kolmogorov-Smirnov test, estimating its associated mean and standard deviation. Finally, the p-value was obtained by evaluating the surrogated cumulative distribution at the sGC original value.

Note that an optimal strategy would be to generate a surrogate data set with the same GC as the original time series but with balanced excitatory and inhibitory interactions. As this situation is unpracticable without knowing the specific anatomical structure of the recordings, what we tested was the error induced in the estimation of the *A* coefficients with signals of common properties, with the same amplitude and frequency as the original signals but without a G-causal link.

**2.4- Neural motifs**

To test the validity of our approach, we simulated several neuronal motifs with different structural connections (see figure 1). The motifs were composed of three to four nodes connected by chemical synapses, which were either excitatory or inhibitory. Each node was a neuronal population composed of 400 excitatory and 100 inhibitory neurons described by the Izhikevich model[53], which receives 50 synapses (sparse connectivity 10%) from randomly selected (excitatory or inhibitory) neighbours in the same population. These parameters were based on previous studies that have simulated spiking neuronal population models[41,53,54]. In addition, in all the simulations each neuron in the network received an independent Poisson input of excitatory synapses. For the coupling between the different nodes, in order to produce each different motif, we assumed that each postsynaptic neuron in a receiver population received 20 synapses from randomly selected presynaptic neurons in the sender population.

The analysed time series corresponded to the mean membrane potentials of each population, which were calculated as the average value of the membrane potential $v$ of all neurons within the population. The averaged value of $v$ can be thought of as a crude approximation of a LFP recording.

The membrane potential $v$ and the recovery variable $u$ of each neuron are described by[53]:

$$\begin{aligned} \frac{dv}{dt} &= 0.04v^2 + 5v + 140 - u + \sum I_x \\ \frac{du}{dt} &= a(bv - u) \end{aligned} \quad (12)$$

The summation $\sum I_x$ is over all the synaptic currents. The model establishes that when $v \geq 30$ mV, its value is reset to $c$, and $u$ is reset to $u + d$. The parameters $a, b, c$ and $d$ determine the firing pattern of the neuron. We employed $(a, b) = (0.02, 0.2)$ and $(c, d) = (-65, 8) + (15, -6)\sigma^2$ for excitatory neurons and $(a, b) = (0.02, 0.25) + (0.08, -0.05)\sigma$ and $(c, d) = (-65, 2)$ for inhibitory neurons, where $\sigma$ is a random variable uniformly distributed on the interval [0,1] that determines the proportion of different spiking neurons (between regular spiking to bursting modes).

In particular, to test the robustness of sGC against neuronal variability the parameters $c$ and $d$ in equation 12 were re redefined as:

$$\begin{aligned} c &= -55 - x + (5 + x)\sigma_1^2 - (10 - x)\sigma_2^2 \\ d &= 4 + y - (2 + y)\sigma_1^2 + (4 - y)\sigma_2^2 \end{aligned} \quad (13)$$

Both $\sigma_1$ and $\sigma_2$ were random variables uniformly distributed in the interval [0,1]. We simultaneously varied $x$ and $y$, keeping the relation $y = 2x/5$ in order to change the proportion of different types of excitatory neurons. Therefore, the maximum values of $c$ and $d$ varied along the line $d = -\frac{6c}{15} - 18$. For example, when $c_{max} = -55$ (which occurs for $x = 0$) most neurons were intrinsically bursting neurons, but there were also regular spiking and chattering neurons.

The synaptic current $I_x$, which can be excitatory, mediated by AMPA ($A$), or and inhibitory, mediated by $GABA_A$ ($G$) is described by the following equations:

$$I_x = g_x r (E_x - v) \tag{14}$$

where $x = A, G$ and $E_A = 0$ mV and $E_G = -65$ mV are the reversal potentials. Unless otherwise stated, all excitatory (inhibitory) weights were set to $g_A = 0.5$ nS ($g_G = 2$ nS). The dynamics of the fraction of bound synaptic receptors $r_x$ is given by:

$$\tau_x \frac{dr_x}{dt} = -r_x + D \sum_k \delta(t - t_k) \tag{15}$$

The summation over $k$ stands for pre-synaptic neurons. D was taken, without loss of generality, equal to 0.05. The time decays were taken as $\tau_A = 5.26$ ms and $\tau_G = 5.6$ ms. Each neuron was subject to an independent Poisson input, representing $n$ pre-synaptic neurons, with a spiking rate $R/n$. Unless otherwise stated $R = 600$ Hz. The Poissonian synapses were assumed as excitatory (AMPA) connections. For each set of parameters, we ran the simulation for 24 seconds with a sample rate of $F_s = 20$ kHz. For the GC analysis, we downsampled the data at 250 Hz. We discarded the first 4 seconds to eliminate transient states, until the signal had a *covariance stationary*, i.e. with constant mean and variance.

In order to study the effects of plasticity on the sGC measures, we incorporated a hybrid spike-timing dependent plasticity (STDP) rule[54,55] in the simulations. To this aim, the excitatory conductance of each inter-population synapse was updated as follows:

$$\begin{aligned} g &= g + A_+ exp(-t/\tau_+), \quad \text{if } t > 0 \text{ (additive LTP)} \\ g &= g - A_- exp(t/\tau_-), \quad \text{if } t > 0 \text{ (multiplicative LTD)} \end{aligned} \tag{16}$$

where $\tau_- = \tau_+ = 5$ ms, $A_- = 1.0$ and $A_+ = 0.5$ nS. For each set of parameters with plasticity, we ran the simulation over 48 seconds.

The effect of the measurement noise was modelled by varying the signal to noise ratio (SNR). For every signal, the SNR was computed as $SNR = 20 \log_{10} A_x/A_n$, where $A_x$ is the mean amplitude of the extracellular membrane potential of one population (renamed as $X(t)$) and $A_n$ the mean amplitude of the noise. Noisy temporal series ($N(t)$) were constructed matching amplitude spectrum and signal distribution with $X(t)$ [56]. The amplitude of the noisy time-series was then scaled as a function of the desired SNR and added to the original signal: $X_n(t) = X(t) + kN(t)$, where $k = 10^{SNR/20}$.

### 2.5- Electrophysiological recordings

To test the proposed methodology in real brain signals, we used a dataset of hippocampal electrophysiological recordings in rats[40]. Briefly, a total of N = 5 animals were implanted with a multichannel electrode along the dorsal hippocampus. For each subject, we selected three LFP signals, recorded at the *stratum radiatum* and *stratum lacunosum-moleculare* in CA1 and one in the dentate gyrus (DG), with the animal freely moving in an open field for 10 minutes. These layers have specific patterns of activity largely contributed by inputs from different afferent pathways to the CA1 region, including the Schaffer collaterals from CA3 to the *str. radiatum*, the layer III of the entorhinal cortex (EC) to the *str. lacunosum-moleculare*[57,58] and the layer II of the EC to the DG through the performant pathway[57]. They were firstly identified using their electrophysiological activity, with the presence of sharp-waves in *str. radiatum* and a maximal theta activity in *str. lacunosum-moleculare* and the inversion of the theta rhythm in the DG. The location of the electrodes was further verified post-mortem with the histological analysis. The data was acquired at 5 KHz, high-pass filtered at 0.5 Hz to remove the continuous component and Notch-filtered at 50 and 100 Hz to remove the net noise and its first harmonic. For the analysis, the data was downsampled at 250 Hz after low-pass filtering at 250 Hz to avoid aliasing.

### 3- Results

### 3.1- Identifying excitatory/inhibitory connections

The first step to validate our approach entailed testing the main assumption underlying sGC, namely, the relationship between the A coefficients of the AR model and the type of projection (excitatory or inhibitory). To do that, we tested 30 motifs composed of three populations (figure 1a). Maintaining a constant population structure, each motif presented a different connectivity pattern, where for each link we selected one of the three options: only excitatory, only inhibitory or no direct link. Therefore, we expected a positive sGC for those links with excitatory projections and a negative sGC for those with inhibitory projections. The connectivity of all motifs was set to encompass many different possibilities (figure 1b), from the simplest case with only one link, to the most complex case, with excitatory and inhibitory connectivity among all populations.

When we applied the sGC, the connectivity was measured by GC analysis (figure 2a; 100% of hits; GC of excitatory links: 0.0864 ± 0.0176; GC of inhibitory links: 0.0244 ± 0.0076; mean ± sd), computing the synaptic ratio only in those networks with a significant G-causality. In summary, 88 links were analysed (48 excitatory, 40 inhibitory). The results show that sGC differentiates both conditions in all cases (excitatory sGC: 0.926 ± 0.026; inhibitory sGC: -0.746 ± 0.145; mean ± sd; $p<0.05$), which confirms its ability to identify the type of synapses (excitatory or inhibitory) at every connection. Moreover, different window lengths (see Methods) were used to test the robustness of the method, including 2.5, 5 and 10 seconds (results not shown). In all cases, 100% of the links were correctly classified as excitatory or inhibitory.

Imposing the zero constraints and fixing the model order at p=15 (60 ms), a total of $A = K^{2p} = 135$ coefficients were computed for each motif. The number of elements eliminated by the Bottom-Up strategy was 7.83 ± 2.32 (mean ± sd), while the Top-Down imposed a total of 69.30 ± 5.01 (mean ± sd) constraints. Combining both strategies, the percentage of information deleted of the whole A matrix was 13.35% ± 1.72% (mean ± sd), measured as $100(1 - (\bar{A}_{cons}/\bar{A})$, where $\bar{A}_{cons}$ and $\bar{A}$ were the mean of the modulus of all A coefficient with and without constraints, respectively. The sGC was also computed using the original A matrix, revealing significant lower values as compared to the restricted version ($p<0.0001$, paired t-test, results not shown), although finding the correct mechanism (excitation or inhibition) in all cases.

**3.2- Determining the excitatory/inhibitory ratio**

We also studied a second type of motif with four nodes. The results of the sGC again inferred the correct G-causality matrix (figure 3), also identifying the sign of the links in those cases where all the projections were only excitatory or inhibitory. The input from population 2 to population 4 had both types of synapses and, although this link was found, the sCG showed a high positive ratio, suggesting a prevalence of excitation over inhibition.

We then created a modified version of this model by increasing the inhibition in the 2→4 connection to characterize the ratio of synapses when both types of projection were present in the same connection. As the main activity in the original motif was excitatory, the weight of the inhibition was progressively increased while keeping the rest of the parameters constant. The results showed a reduction of the GC which matched a negative trend in sGC when the inhibition overcame the excitation (figure 4). The functional connectivity was restored when the negative sGC dominated the system, increasing systematically with higher values of inhibitory conductance. The correlation between the sGC and the inhibitory conductance followed a sigmoidal function, reaching values near the maximum and minimum of the sGC when the inhibition was lower than 2 nS and higher than 4.5 nS, respectively. The intermediate window can be approximated to a linear regression of both factors ($\rho^2 = 0.92$; p<0.0001; slope b = -0.75). In our case, the balance was found for an inhibition weight $g_G \sim 3.25$ nS and $g_A = 0.5$ nS. Thus, as the inhibition increased, the sGC rate became more negative. The projections were identified as predominantly inhibitory for $g_G > 5$ nS (sGC = -0.91).

In order to study the robustness of our method against the parameters of the model, we compared the connectivity patterns obtained for the four nodes motif in five extra cases. First, we performed simulations in the presence of plasticity rules for the excitatory synapses inter-populations (see Methods for more details). In figure 5a we compared how sGC changed with the inhibitory conductance from node 2 to node 4 when STDP rules were applied. The results show a low dependence of the sGC against plasticity, without significant differences in most of the cases compared to the original motif. Only when node 2 was highly inhibiting node 4, the backward link (4→2) was not properly identified. Secondly, the robustness of the method against measurement noise was tested by changing the SNR (figure 5b), from the perfect situation (without added noise) to SNR = -1 (see Methods). Both the GC and the sGC exhibited proportional decays with increasing noise level, remaining the sign of the sGC stable for all SNR values. Even in conditions of high noise, if the temporal GC was able to detect the functional link, the sGC was also able to reflect the excitation and inhibition projections. Third, for fixed synaptic conductance, we changed the amount of internal variability of the neuronal cells in each population, by modifying the proportion of neurons with different firing patterns (figure 5c). We observed that sGC reflected the expected functional connectivity of the network despite the change in the ratio between regular spiking and bursting neurons. Fourth, we changed the amount of external noise received by each neuron, which altered the amount of averaged activity in the network. In figure 5d we plot sGC as a function of the Poisson rate. The G-causal network estimated by the method is invariant to different noise inputs in three links, with a sGC close to 1 or -1 from node 2 to node 4, for those inhibitory conductances that allowed for a clear distinction between excitatory or inhibitory sGC in the original model (figure 4). Five, the effect of the sampling rate was tested by replicating the analysis and changing the sampling frequency of the data (figure 5e and 5f). The sGC ratio decayed progressively as a function of the sampling rate, matching with the results of the GC. Even if the information included in the AR was similar (the model order in ms was approximately constant, figure 5f), a higher sampling frequency implies a larger number of parameters to estimate, therefore increasing the errors in the measurements. These results validated the use of sGC in all those conditions where the GC can be applied.

### 3.3- Positive and negative interactions are present in the hippocampus

We tested whether the proposed methodology offers new information about the functional connectivity in the well-known hippocampal circuit. We selected two LFP signals from the rat CA1 region, recorded at the *str. radiatum* (RAD) and *lacunosum-moleculare* (LM) and a third LFP recorded at the DG (figure 6a and 6b). The results of GC showed a significant functional connectivity in all directions, being higher in the direction from LM to RAD (figure 6c). In contrast, the sGC revealed distinct couplings, positive in both links to RAD, negative and positive from RAD to LM and from DG to LM, respectively, and negative from LM to DG, while there is no clear sign from RAD to DG. These results were highly consistent across subjects and resistant to variations in the sliding time window length (between 5 and 20 s) and model order between 12 and 24 (48 and 96 ms). Interestingly, the largest values of sGC (negative from RAD to LM and positive from DG to LM) were not associated to the highest GC. We also found that the link with the strongest connectivity measured by GC was associated with a modest and positive sGC. We repeated the analysis of sGC without following the proposed strategies, that is, using the original A matrix without imposing zero constraints, finding similar results. However, the absolute value of the sGC was lower compared to the analysis with restricted coefficients ($p<0.05$, paired t-test, results not shown). These results confirmed that our methodology outperforms classic algorithms estimating AR models which do not impose constraints to the computed coefficients.

## 4- Discussion

While the GC is often interpreted as a tool to detect the information flow in terms of bit rate[30,59], it remains unclear the functional meaning of that information. Here, we presented a new GC-based method aimed at inferring correlations on Granger-causal interactions of brain signals. In this way, we are able to better characterize the connectivity of the network by determining whether a GC is positive or negative, and relating this result with two different functional mechanisms. We successfully tested it in a specific paradigm, where excitation and inhibition are supposed to interact inversely in a G-causal link, generating different coefficients in the AR model. These two scenarios (i.e. excitation and inhibition) have quite different roles in brain networks[38,39], but their estimation is not possible when GC is applied. Our approach demonstrates that the GC of a particular link may be positive or negative, and its sign should be associated to different functional meanings. It is important to note that the specific meaning will depend on the signals being used. This is, positive or negative GC coefficients will likely reflect different underlying neurobiological mechanism when obtained in the analysis of blood oxygenation level dependent (BOLD) signals from fMRI experiments, or when obtained from intracranial electrophysiological recordings. But in all cases, however, positive and negative coefficients will highlight fundamentally different statistical interdependencies in the signals.

The effect of excitatory and inhibitory neurons in functional connectivity has been studied using multi-unit spike activity from large-scale neuronal networks[60]. Furthermore, previous works have demonstrated the utility of AR models to observe positive and negative couplings in fMRI[61]. This methodology generally estimates an order-1 model to compute the GC, identifying "excitatory" or "inhibitory" pathways based on the sign of the only coefficient extracted. However, several limitations, e.g., the invariable model order[62], have prevented their application to datasets with quite diverse time scales. Our approach overcomes these issues, as it considers all the significant coefficients without prior restrictions of model order, proposing a generic methodology that can be extrapolated to every GC study. Moreover, we have proposed two strategies to mitigate the effect of artefactual parameters included in the AR models, by imposing zero constraints to those elements that are not significantly contributing to the prediction. While these steps are not essential to determine the sGC value and other approaches can be used instead[28], the results were always improved when observed (i.e. the absolute value of sGC was higher compared to the non-restricted model).

Our model revealed that excitation and inhibition contribute differently to the measured connectivity. In a balanced situation, the G-causal influence of an excitatory link showed the highest values (figure 1b). Furthermore, when both excitation and inhibition are present in a link, they compete to control the target population. Our analysis suggests that, for the neuronal dynamics and composition of neural networks used in our models, excitatory activity tends to mask the inhibitory one, as the sGC ratio did not change significantly as compared to the case when only excitatory projections were present (figure 3c). These results demonstrate a correlation between the degree of GC and the type of synapse. However, an increase in the weight of the inhibitory projections progressively reduces the sGC ratio, until breaking the balance and changing the sign of the sGC (figure 4). This might provide valuable information on the dynamics of brain connections in which the influence of one region over its efferent targets varies dynamically with, or as a consequence of, the cognitive needs.

The analysis of real LFP signals using sGC unveiled different couplings in the directed functional connectivity of the hippocampus. We computed a multivariate AR model using three signals recorded at specific layers of the CA1 region and the dentate gyrus. However, some limitations of the signals need to be mentioned before speculating about the neurobiological meaning of the obtained sGC results. First, it must be considered that LFP signals are affected by volume conduction that may create spurious correlations between recordings[63]. Second, the multivariate AR model allows the discrimination of direct and indirect links between the signals included in the model as long as all the relevant links are represented. Therefore, this limitations needs to be considered when applying these methods to a particular dataset[41]. Moreover, the GC is also sensitive to the signal-to-noise ratio of both time-series[64], thus, the high theta power in LM may be affecting the prediction on RAD and DG. With these limitations in mind, we interpret the sGC results as follows: the activity reaching the distal part of the dendrites of CA1 pyramidal neurons from the EC (LM) positively interacts with the inputs arriving downstream in the dendrite from CA3 (RAD). This positive interaction may translate into facilitated integration of afferent information from both pathways. Contrary, the negative interaction found from RAD to LM, may reflect competition between the two inputs and the segregation of the transmission channels. The mechanism underling this negative interaction could be the activation of oriens lacunosum-moleculare (OLM) interneurons, known to receive inputs from the Schaffer collateral and project to the LM inhibiting the EC input[65]. Therefore, the balance between integration and segregation could be dynamically controlled by the inhibitory tone. The positive sGC from the DG to RAD may represent the information flow through CA3, following the trisynaptic pathway in the hippocampus. Regarding the interaction between LM and DG, it can be explained by the connectivity in the EC since layers II and III in this neocortical structure project to the DG and LM, respectively. The negative sGC from LM to DG may accounting for the inversion of the hippocampal theta rhythm between CA1 and the DG[7]. Finally, there is not a clear sign in the sGC from RAD to DG which also exhibits the lowest value of connectivity, in consonance with the lack of a direct anatomical pathways between these structures.

Note that GC and sGC are complementary indices and can be simultaneously computed. The former estimates the amount of information transmitted (information flow), while the latter represents, in our framework, a closer look to the underlying causal mechanism[3,30]. Although the concept of *causality* is embedded in GC, it implies statistical dependences rather than true physical causal interactions among the variables[30]. In this line, GC and other methods based in AR models have been classified as *directed functional connectivity* indexes, while the term *effective connectivity* is used when determining the circuitry that causes an effect in neural systems. In this line, there are other approaches, like DCM[29], which find the optimal mechanism behind the functional links. We have compared the GC with the sGC for simplicity, as the formulation of the sGC derives from the same AR models and shares the same limitations. However, it can be computed jointly with other statistical connectivity indexes like PDC, TE or PhTE. For instance, the pairwise spectral connectivity could be firstly estimated by other metrics, like PDC, and then positive or negative relationship at a specific frequency by filtering the signal and computing the sGC.

In this work, we have stablished the basis of the sGC in the temporal domain and under the assumption of stationarity. It will be interesting to test the performance of sGC in other conditions. For instance, it can be easily extended to a time-varying version with a windowing methodology[32,49]. In this case, the signal could divided into small epochs considered as stationary, and then the AR model would be computed for each time window[49]. Future studies include the characterization of positive and negative relationships as a function of the frequency, thus combining the frequency and time varying approaches. An important step will be to separate the elements of the AR model used to compute the GC and the sGC. In the former, it is the variance of the residuals what determines the connectivity, while only some coefficients of the AR model are required to compute the sGC. Following the Fourier transform, the AR model can be expressed by a transfer function ($H(w)$) and an error[66]. The information of the A matrix that determines the sGC is, therefore, in the transfer function. The analysis of both, the real and imaginary part of the complex values of these elements, shall determine the spectral sGC.

We have tested this proposal in a theoretical paradigm to demonstrate its efficacy, but several aspects need to be considered to prevent misinterpretations. First, in order to be interpretable, the polarity of the used signals needs to be known. The polarity of the simulated currents in our model is known, yet it is more difficult to establish in most of the LFP and EEG recordings[63,67]. Therefore, the sign of the GC in these datasets might not be directly interpreted as a readout of excitation and inhibition. Secondly, some conditions such as signal non-stationarity and non-linearity were not considered in this work but might be present in biological signals. Recent implementations of GC overcome these issues[49,68,69], opening the possibility to implement sGC concepts on these indices. Thirdly, electrophysiological recordings are generated by the contribution of many sources spatiotemporally overlapped[63,70,71]. Thus, prior knowledge of the system and signals is required. Overall, a positive GC might be interpreted as a directed coupling in which the involved populations, sender and receiver, are linked in a comparable functional state. A negative GC would, complementary, reflect a link between populations in distinct functional states.

## 5- Author Contributions

V.L.M., C.M., E.P. and S.C. conceived the project. V.L.M. and F.M. carried out the research. E.P. checked the mathematical methods and did the statistical analysis. All the authors discussed the results. V.L.M. wrote the manuscript and all the authors reviewed it.

**6- Competing Interests**

The authors declare having no competing interests.

**7- Acknowledgments**

We thank Elena Pérez-Montoyo for her assistance with the electrophysiological datasets and Dr. Daniele Marinazzo and Dr. Simona Obreja for providing useful comments on the manuscript. This work was supported by the Spanish Ministry of Economy and Competitiveness (Project Nos. TEC2016-800    63-C3-3-R, TEC2016-80063-C3-2-R, and BFU2015-64380-C2-1-R), the European Union through the Horizon 2020 programme (Grant Agreement No. 668863, SyBil-AA), the Spanish State Research Agency (SEV- 2013-0317), the Brazilian Coordenação de Aperfeiçoamento de Pessoal de Nível Superior (Grant Nos. 88881.120309/2016-01 and 88881.068077/2014-01), the Brazilian Conselho Nacional de Desenvolvimento Cientìfico e Tecnològico (Grant No. 432429/2016-6) and the US National Science Foundation (grant NSF IIS-1515022). V.J.L. was supported by a predoctoral fellowship La Caixa-Severo Ochoa from Obra Social La Caixa. SC and CM acknowledge the Spanish State Research Agency, through the Severo Ochoa and María de Maeztu Program for Centers and Units of Excellence in R&D (SEV-2017-0723 and MDM-2017-0711, respectively).

# 8- Bibliography

**Figures**

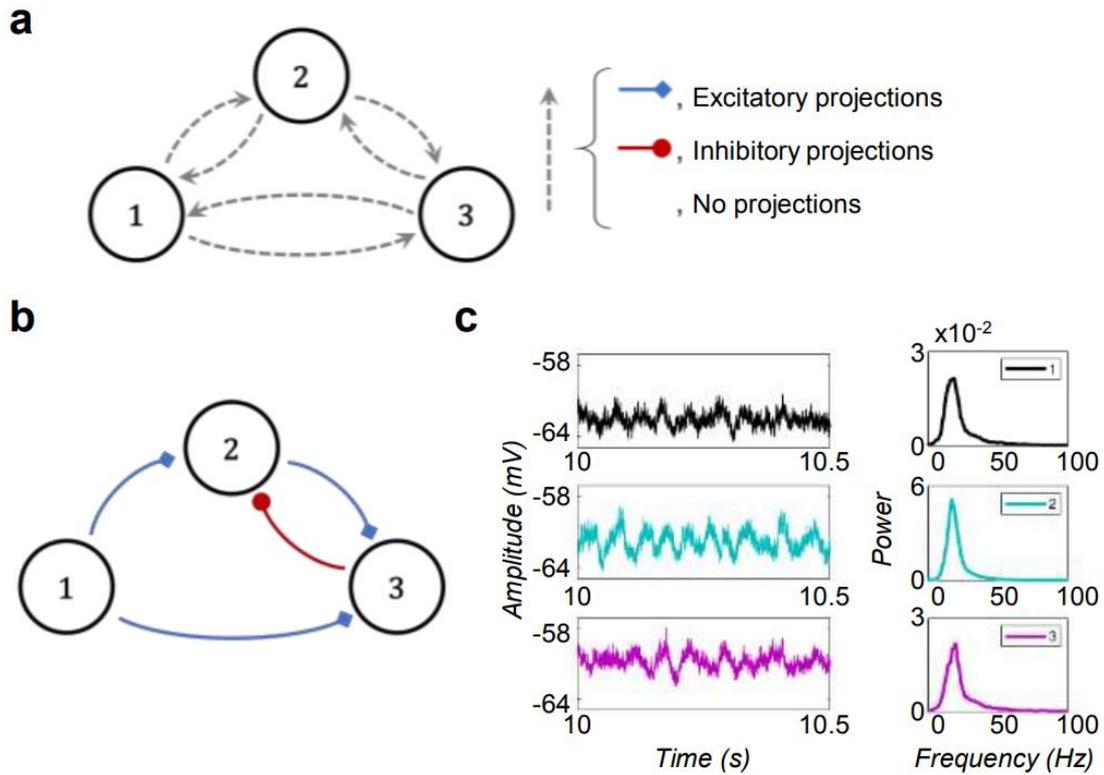

Figure 1: (a) Scheme of the structure of all implemented motifs. Broken arrows represent the different options for interconnecting the populations that change for each motif. Links can be excitatory, inhibitory or non-existent. (b) Representative example of one motif, including three links with excitatory projections (blue arrows) and one with an inhibitory projection (red arrow). (c) Examples of the temporal evolution of the mean membrane potential for the three populations in B and their power.

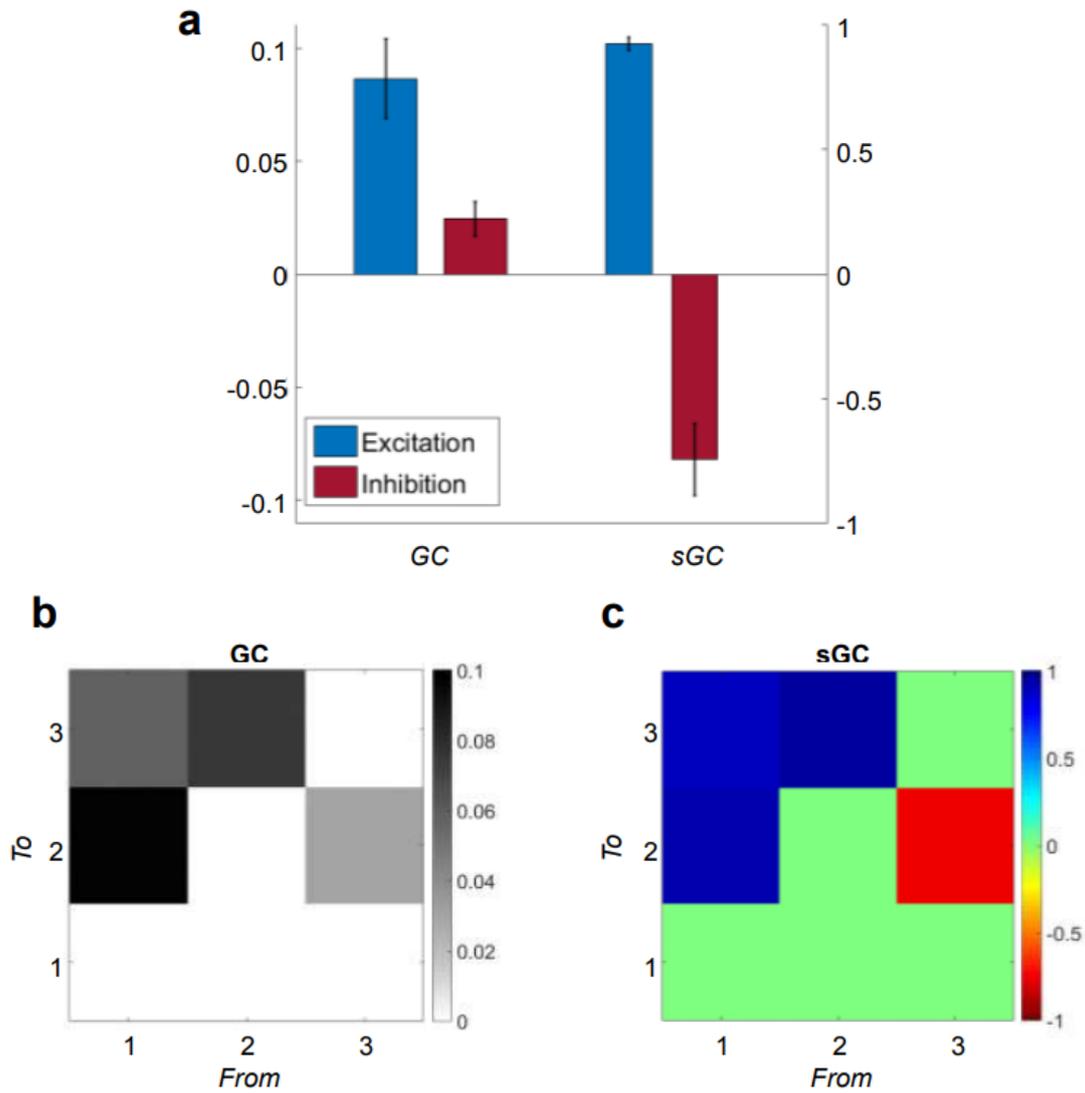

Figure 2: (a) Average of GC and sGC across links, revealing differences in both conditions. Links from an excitatory node exhibit a higher GC and sGC value than those from an inhibitory one. (b) Representative GC matrix for the motif showed in figure 1B. All significant values correspond to the implemented structure. (c) sGC ratios for each bind in figure 1B, where positive and negative values match excitatory and inhibitory connectivity, respectively.

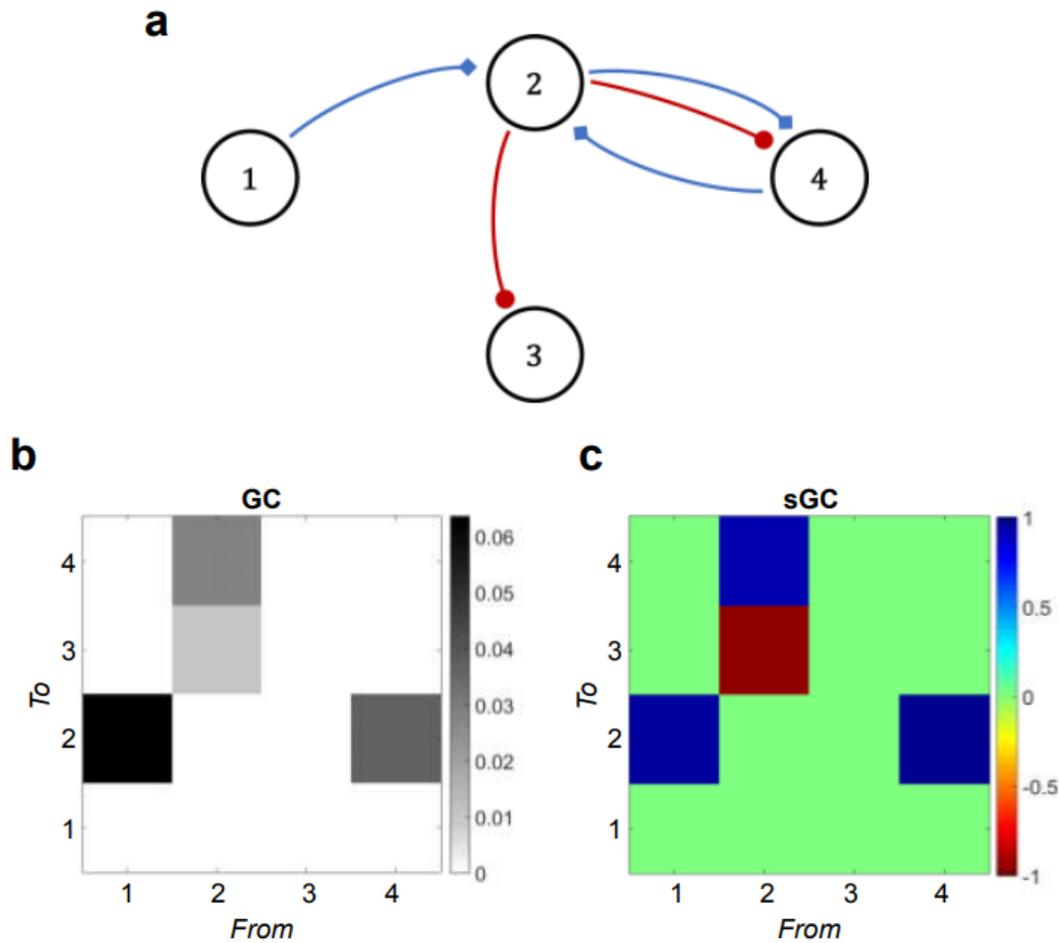

Figure 3: (a) Scheme of the analyzed motif. Red and blue arrows indicate inhibitory and excitatory projections, respectively. (b) Resulting matrix of GC. Nonzero values represent links with significant connectivity, which match with the physical network. (c) Matrix with the sGC ratio. Only values with significant sGC are represented (p<0.05). Values near 1 represent predominant excitation; those close to -1 represent higher inhibition. Pairs with a single type of projection (excitatory or inhibitory) are correctly identified, although the projections from population 2 to 4 are considered excitatory, suggesting that excitation is monopolizing the functional connectivity.

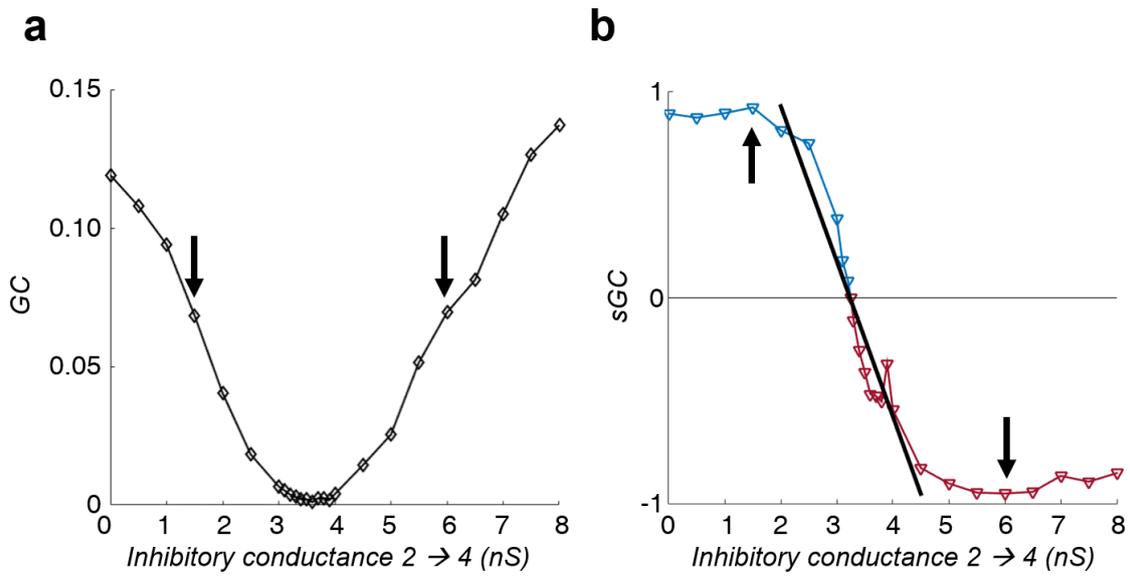

Figure 4: (a) GC is highly related to the combination between types of projection. For a fixed value of excitation, a moderate increase of the inhibitory conductance (between 3 and 4 nS) suppresses the functional connectivity, while for higher values (from 5 to 8 nS) it raises again, likely reflecting a competition between excitation and inhibition in the link. (b) The sGC ratio reveals the balance between excitation and inhibition, with a working zone of linear correlation (from 2 nS to 4.5 nS). The progressive increase of the inhibition counterbalances the excitation and minimizes the GC, until it dominates the link (from 5 to 8 nS) and enhances the functional connectivity. Note how for common values of GC (black arrows), the sGC perfectly differentiates both conditions.

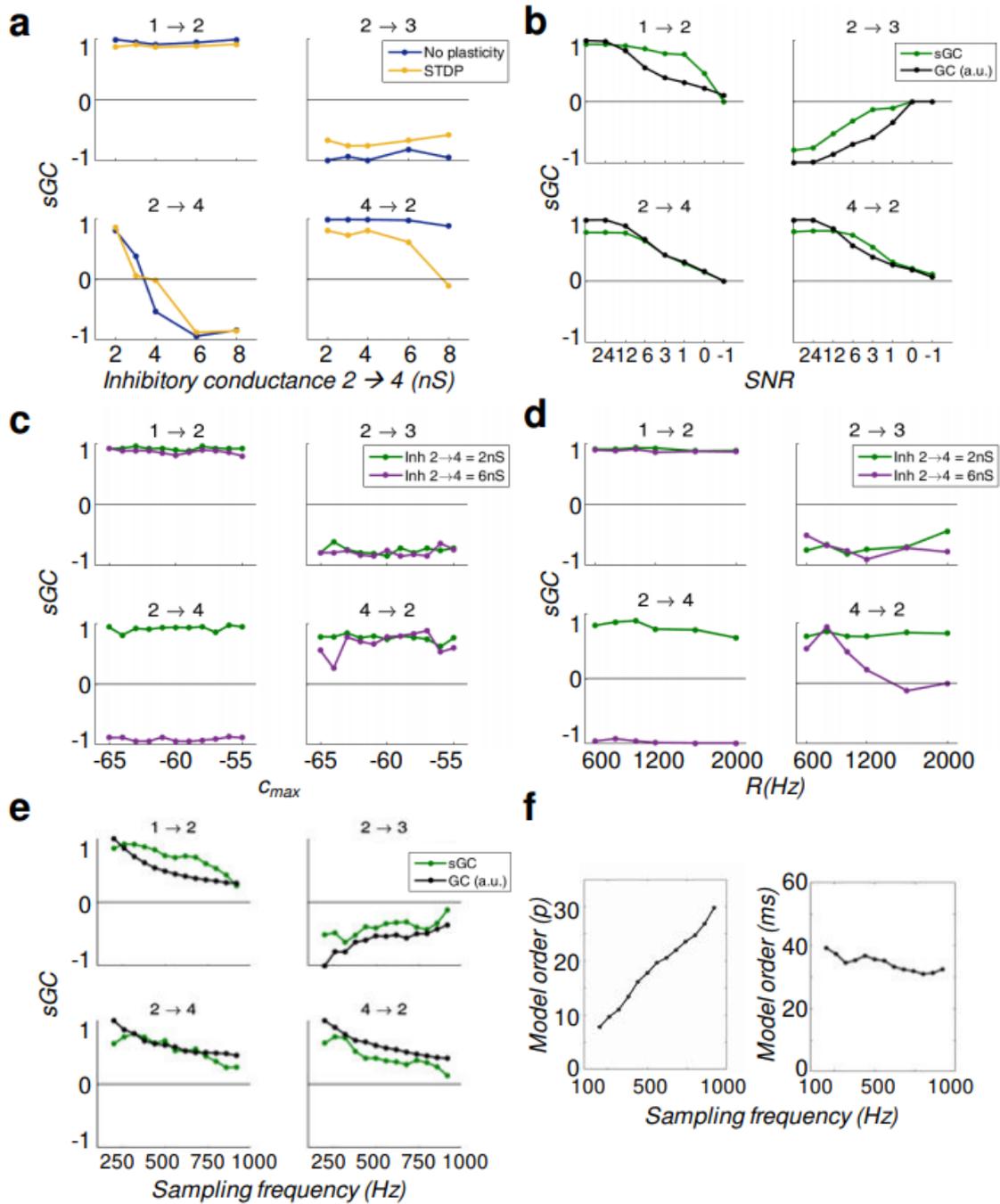

Figure 5: (a) Evolution of sGC ratio in each link of the motif, with and without STDP plasticity rules (yellow and blue lines, respectively), as a function of the inhibition weight from population 2 to 4. Changes in the value of this projection only affect the sGC results in that link, while the other links remain the same. An increase in the inhibition weight, involving a change in the excitatory/inhibitory ratio, is captured by the method, showing opposite ratios for inhibition weights of 2 nS and 8 nS (where the excitation is kept constant). (b) Variation of GC and sGC in function of the SNR. Both measurements present similar decays when the external noise is increased, remaining the sign of the sGC

(excitation/inhibition) stable even for low SNR. GC values were normalized by setting the maximum value to 1, and GC corresponding to link 2→4 has been inverted for comparison purposes with sGC. (c) sGC in function of the internal variability of the population, for a case with low (green) and high (purple) inhibitory conductance from 2 to 4. In all cases, the sGC remains stable. (d) Variation of sGC for different noise inputs. The method finds similar results in almost every link when the Poissonian rate is increased, except for the connectivity from 4 to 2, which loses accuracy when the opposite projections are predominantly inhibitory. (e) Comparison of GC and sGC for different sampling rates. The GC decays for higher values of sampling frequency, as well as the sGC. Nevertheless, the sGC finds the correct coupling in all cases. (f) Model orders estimated using the AIC for the simulations in e, with their values in samples (left) and in milliseconds (right).

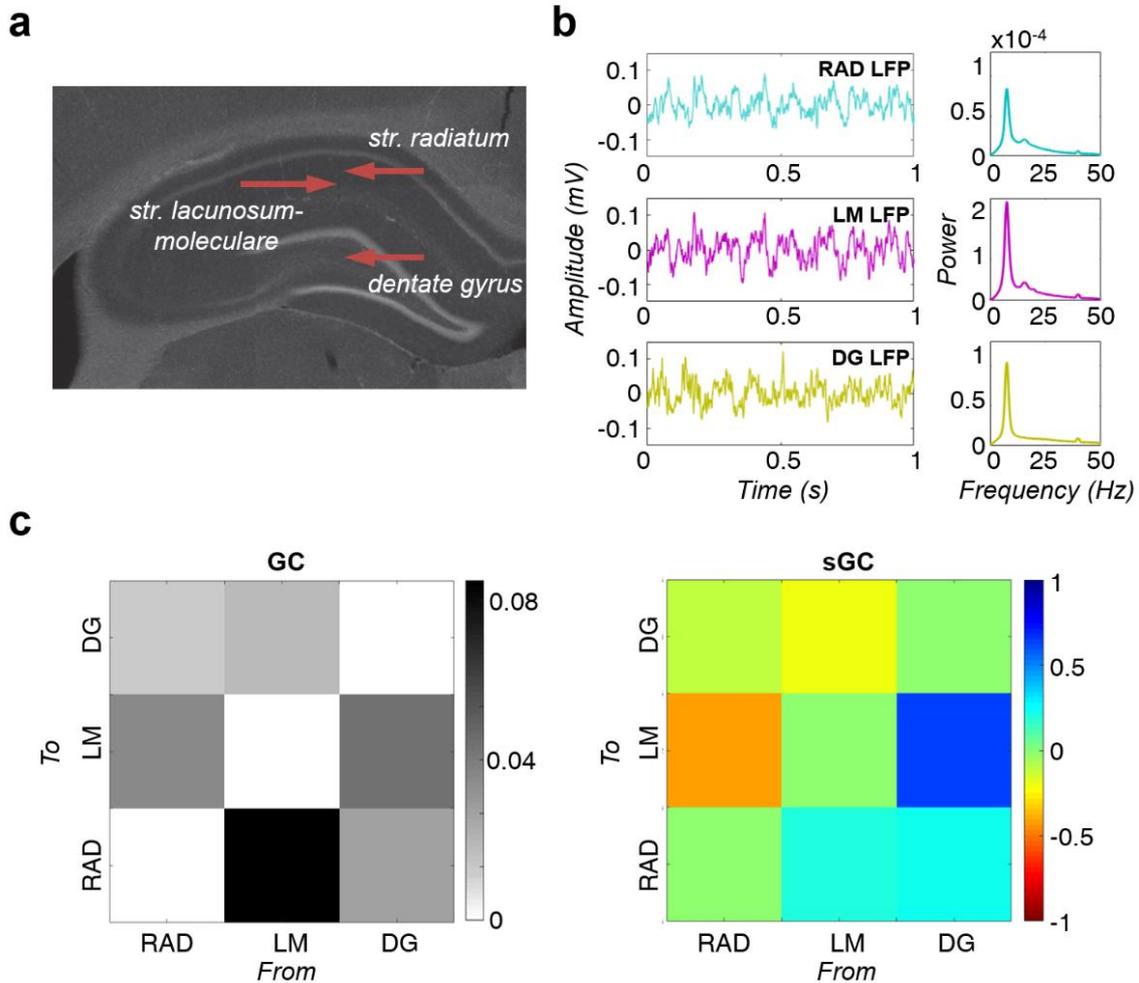

Figure 6: (a) Histological recording of the hippocampus showing the location of the stratums recorded. (b) Example of time-series and power spectrum of the LFPs recorded at *stratum radiatum* (RAD LFP), *stratum lacunosum-moleculare* (LM LFP) and at the dentate gyrus (DG LFP). (c) Results of GC (left) and sGC (right) between hippocampal LFPs. The strongest connectivity is found from LM to RAD, which correspond to a positive sGC with the same value than from DG to RAD. while the links to LM present the highest values of sGC (negative and positive from RAD and from DG, respectively).